\documentclass[pra,twocolumn,showpacs,superscriptaddress,amsmath,amssymb]{revtex4}
\usepackage{graphicx}

\usepackage{color}

\begin{document}
\title{ Gauge symmetry in Kitaev-type spin models and index
theorems on odd manifolds}
\author{Yue Yu}
\affiliation{Institute of Theoretical Physics, Chinese Academy of
Sciences, P.O. Box 2735, Beijing 100080, China}
\date{\today}
\begin{abstract}
We construct an exactly soluble spin-$\frac{1}2$ model on a
honeycomb lattice, which is a generalization of Kitaev model. The
topological phases of the system are analyzed by study of the
ground state sector of this model, the vortex-free states.
Basically, there are two phases, A phase and B phase. The
behaviors of both A and B phases may be studied by mapping the
ground state sector into a general $p$-wave paired states of
spinless fermions with tunable pairing parameters on a square
lattice. In this $p$-wave paired state theory, the A phase is
shown to be the strong paired phase, an insulating phase. The B
phase may be either gapped or gapless determined by the
generalized inversion symmetry is broken or not. The gapped B is
the weak pairing phase described by either the Moore-Read Pfaffian
state of the spinless fermions or anti-Pfaffian state of holes
depending on the sign of the next nearest neighbor hopping
amplitude. A phase transition between Pfaffian and anti-Pfaffian
states are found in the gapped B phase. Furthermore, we show that
there is a hidden SU(2) gauge symmetry in our model. In the gapped
B phase, the ground state has a non-trivial topological number,
the spectral first Chern number or the chiral central charge,
which reflects the chiral anomaly of the edge state. We proved
that the topological number is identified to the reduced
eta-invariant and this anomaly may be cancelled by a bulk
Wess-Zumino term of SO(3) group through an index theorem in 2+1
dimensions.

\end{abstract}


 \maketitle

\section{Introduction}\label{secI}
The concept of the topological order recently is widely
interesting the condensed matter physicists because it may
describe the different 'phases' without breaking any global
continuous symmetry of the system \cite{wenniu}. However, unlike
the conventional order related to the symmetry of the system in
Landau's phase transition theory, the topological order of quantum
states is not well defined yet. For example, in the quantum Hall
effects, the topological property of the quantum stats may be
reflected by the filling factor of the Landau level which may be
thought as a topological index, the first Chern number in magnetic
Brilliouin zone \cite{ttnn,niu}. Nevertheless, only the first
Chern number can not fully score the topological order of the
quantum Hall states. In a given filling factor, the quasiparticles
may obey either abelian or non-abelian statistics. On the other
hand, the edge state may partially image the topological
properties of the bulk state \cite{wen}. In quantum Hall system,
it was seen that the edge state may be described by a conformal
field theory \cite{mr}. Thus, according to the bulk-edge
correspondence due to the gauge invariance, it shows that the bulk
state is determined by a Chern-Simons topological field theory
\cite{cs}. However, the bridge between the microscopic theory of
the two-dimensional electron gas and the Chern-Simons theory was
not spanned.

Kitaev recently constructed an exactly soluble spin model in a
honeycomb lattice \cite{ki}. Using a Majorana fermion
representation, he found the quantum state space is characterized
by two different topological phases even there is not any global
symmetry breaking. The A phase is a gapped phase which has a zero
spectral Chern number and the vortex excitations obey abelian
anyonic statistics. The B phase is gapless at special points of
Brilliouin zone. When the B phase is gapped by a perturbation, it
is topologically non-trivial and has an odd-integer spectral Chern
number. (We call the gapless B phase the B1 phase and gapped one
the B2 phase.) Kitaev showed that if the spectral Chern number is
odd, there must be unpaired Majorana fermions and then the vortex
excitations obey non-abelian statistics. Consistent with the
non-abelian statistics, the fusion rules of the superselection
sectors of Kitaev model are the same as those of the Ising model.
However, the source of the non-abelian physics has not been
clearly revealed yet. On the other hand, the first Chern number
can only relate to an abelian group and therefore, an odd spectral
Chern number leads to a non-abelian physics but an even one did
not is topologically hard to be understood.

Although Kitaev model has a very special spin coupling, its very
attractive properties caused a bunch of recent studies
\cite{note,bas,duan,cn,yw,yzs,yu,lzx,yk,sdv,lkc,sdm,kw}. It is
convenient to understand Kitaev model if one can map this model to
a familiar model. In fact, Kitaev model may be mapped into a
special $p$-wave paired BCS state if only the vortex-free sector
of the model is considered \cite{cn}. We recently generalized
Kitaev model to an exactly soluble model whose vortex-free part is
equivalent to
$\Delta_{1x}p_x+\Delta_{1y}p_y+i(\Delta_{2x}p_x+\Delta_{2y}p_y)$-wave
paired fermion states with tunable pairing order parameters
$\Delta_{ab}$ on a square lattice. \cite{yw}. The phase diagram of
our model has the same shape as that of Kitaev model, i.e, the
boundary of the A-B phases are corresponding to the points ${\bf
p}=(0,0),(0,\pm \pi)$ and $(\pm \pi,0)$ in the first Brilliouin
zone. The A phase is gapped and may be identified as the strong
paring phase of the $p$-wave paired state \cite{rg}. The B phase
can be either gapped or gapless even if T-symmetry is broken. We
find that gapless excitations in the B phase, i.e., the B1 phase,
is protected by a generalized inversion (G-inversion) symmetry
under $p_x\leftrightarrow {\Delta_{1y}\over \Delta_{1x}} p_y$ and
the emergence of a gapped B(B2) phase is thus tied to G-inversion
symmetry breaking. For instance, the $p_x+ip_y$ wave paired state
is gapped while $p_y+ip_y$-wave paired state is gapless although
they both break the T-symmetry. The critical states of the A-B
phase transition remains gapless whether or not T- and G-inversion
symmetries are broken, indicative of its topological nature.
Indeed, if all $\Delta_{ab}$ are tuned to zero, the topological
A-B phase transition is from a band insulator to a free Fermi gas.
The Fermi surface shrinks to a point zero at criticality.

In this paper, we further generalize the model proposed by the
present author and Wang in ref. \cite{yw} to a model whose square
lattice mapping includes a next nearest hopping of the spinless
fermions. In this case, the A phase is still a strong pairing
phase as before. However, the B2 phase has more fruitful
structure. The particle-hole symmetry is broken even if the
chemical potential and the pairing parameters vanish. Near the
long wave length limit (${\bf p}^*=(0,0)$ critical line), the
effective chemical potential has the different sign from that of
the nearest neighbor hopping amplitude. Near other two critical
lines $(0,\pi)$ and $(\pi,0)$, when the next nearest neighbor
amplitude is positive, the effective chemical potential is also
positive. When the next nearest neighbor amplitude is negative,
the effective chemical potential is also negative. A positive
chemical potential corresponds to a closed Fermi surface of the
particles and then a Pfaffian of the particles while a negative
chemical potential to a closed Fermi surface of holes and then an
anti-Pfaffian of the holes of the spinless fermion. Therefore, a
Pfaffian/anti-Pfaffian phase transition happens in the B2 phase.
This Pfaffian/anti-Pfaffian phase transition has been seen in the
context of the $\nu=5/2$ fractional quantum Hall effect
\cite{pfapf1,pfapf2}. The model we present here is exactly the
same as a toy model on square lattice to study the Pfaffian and
anti-Pfaffian physics \cite{pfapf1}. The B1 and B2 phases when the
next nearest neighbor hopping is absent are corresponding to the
particle-hole symmetry is conserved or spontaneously broken.

 The another topic of this paper is trying to reveal the
 mathematical connotation behind the topological order.
 We emphasize that there is a hidden SU(2)
gauge symmetry in this model if the model is represented by
Majorana fermion operators. This non-abelian gauge symmetry is the
source of the non-abelian physics of the model. The non-abelian
degrees of freedom in the A phase are confined while in the B2
phase, the non-abelian degrees of freedom are deconfined. There is
a Wess-Zumino(ZW) term for SU(2)/$Z_2$  group whose lever $k$ may
character the confinement-deconfinement phases. A level $k$ WZ
term corresponds to a level $k$ SU(2)/$Z_2$ Chern-Simons
topological field theory. It was known that $k=1$ theory can only
have abelian anyon while $k=2$ theory includes non-abelian anyons
\cite{witten}. A recently proved index theorem in 2+1 dimensions
 shows that the sum of this WZ term and a reduced
eta-invariant $\bar\eta$ is an integer \cite{daizh}. We show that
difference between the WZ term and a part of the eta-invariant
gives an ambiguity of the WZ term. Another part of this
eta-invariant is identical to the chiral central charge, a half of
the spectral Chern number. Thus, an odd Chern number corresponds
to a $\pi i$ ambiguity while an even Chern number to a $2\pi i$
ambiguity. According to the bulk-edge correspondence, the former
is consistent with $k=2$ while the latter is consistent with
$k=1$.

The rest of this work was organized as follows. In Sec. II, we
recall Kitaev model and show the SU(2) gauge invariance. In Sec.
III, we will describe the generalized model. In Sec. IV, we give
the phase diagram of the system. In Sec. V, we consider the
continuous limit of our model and show that the low energy
effective theory is the Majorana fermions coupled to a SO(3) gauge
field in a pure gauge. In Sec. VI, we apply the index theorem on
odd manifold to our model. In Sec. VII, we present a understanding
to the edge state from the index theorem point of view. The
section VIII is our conclusions. We arrange three appendices.
Appendix A is to address the mathematic expression of the index
theorem on odd manifold because most of physicists are not
familiar with it. In Appendix B, we give an introduction to the
representation to the spin-1/2 in the conventional fermion and
Majorana fermion. And in Appendix C, for completeness, we recall
the vortex excitations in our model although it was studied in our
previous work \cite{yw}.

\section{Kitaev model}

We first recall some basic results of Kitaev model, which is a
spin system on a honeycomb lattice \cite{ki}. The Hamiltonian is
given by
\begin{eqnarray}
H_{ki}=-J_x\sum_{x{-\rm links}}
\sigma^x_i\sigma^x_j-J_y\sum_{y{-\rm links}}
\sigma^y_i\sigma^y_j-J_z\sum_{z{-\rm links}}
\sigma^z_i\sigma^z_j,\nonumber
\end{eqnarray}
where $\sigma^a$ are Pauli matrices and 'x-,y-,z-links' are three
different links starting from a site in even sublattice \cite{ki}.
This model is exactly solvable if one uses a Majorana fermion
representation for spin. Kitaev has shown that his Hamiltonian has
a $Z_2$ gauge symmetry acting by a group element, e.g., for
(123456) being a typical plaque
$$W_P=\sigma^x_1\sigma^y_2\sigma^z_3\sigma^x_4\sigma^y_5\sigma^z_6$$
with $[H_{ki},W_P]=0$. In fact, we can show that this model has an
SU(2) gauge symmetry in the Majorana fermion representation. Let
$b_{x,y,z}$ and $c$ be four kinds of Majorana fermions with
$b_x^2=b_y^2=b_z^2=c^2=1$ and define
\begin{equation}
(\chi^{cd})=\frac{1}2\left(\begin{array}{cc}
b_x-ib_y&b_z-ic\\
b_z+ic&-b_x-ib_y\\
\end{array}\right).
\end{equation}
One observes SU(2) gauge invariant operators
\begin{eqnarray}
{\hat \sigma^a}=\frac{1}2{\rm
Tr}[\chi^\dag\chi(\sigma^a)^T]=\frac{i}2(b_ac-\frac{1}2\epsilon_{abc}b_bb_c)
\end{eqnarray}
with respect to the local gauge transformation $\chi^{cd}\to
U^{cc'}\chi^{c'd}$ and then $(\chi^\dag)^{cd}\to
(\chi^\dag)^{cc'}(U^{-1})^{c'd}$ for $U\in $SU(2) \cite{aff}. It
is easy to check that $\hat\sigma^a/2$ may serve as spin-1/2
operators. Replacing $\sigma^a$ by $\hat\sigma^a$, Kitaev model
has a hidden SU(2) gauge symmetry which is trivial in the spin
operator representation. The constraint $D=1$ is also gauge
invariant because $D=b_xb_yb_zc=-i{\hat\sigma}^x{\hat\sigma}^y
{\hat\sigma}^z$. Under this constraint, $\hat\sigma^a$ takes the
form $ib_ac$ after using $b_xb_yb_zc=1$. The SU(2) symmetry of
$\sigma^a=ib_ac$ can be directly checked
\begin{eqnarray}
\sigma^a=ib_ac=ib_a'c',
\end{eqnarray}
where
\begin{equation}
\left(\begin{array}{c} b_x'\\b_y'\\b_z'\\c'\\ \end{array}\right)
=\left(\begin{array}{cccc}\alpha_1&\alpha_2&\beta_1&-\beta_2\\
                          -\alpha_2&\alpha_1&-\beta_2&-\beta_1\\
                          -\beta_1&\beta_2&\alpha_1&\alpha_2\\
                          \beta_2&\beta_1&-\alpha_2&\alpha_1\\ \end{array}\right)
                          \left(\begin{array}{c} b_x\\b_y\\b_z\\c\\ \end{array}\right)
\end{equation}
with $\alpha_1^2+\alpha_2^2+\beta_1^2+\beta_2^2=1$ and
$b_xb_yb_zc=b_x'b_y'b_z'c'=1$. Using Jordan-Wigner transformation,
a variety of Kitaev model on a brick-wall lattice has been exactly
solved \cite{note} and a real space ground state wave function is
explicitly shown \cite{cn}. This variety should correspond to
another gauge fixed theory.

After some algebras, Kitaev transferred the Hamiltonian to a free
Majorana fermion one \cite{note}
\begin{eqnarray}
H=\frac{1}2\sum_{{\bf p};\mu,\nu=b,w}H({\bf p})_{\mu\nu}c_{-{\bf
p},\mu}c_{{\bf p},\nu}, \label{ham}
\end{eqnarray}
where $H(-{\bf p})=-H({\bf p})$ and $c_{{\bf q},\mu}$ are the
Fourier components of a Majorana fermion operators and $\mu=b$ or
$w$  refers to the even or odd position in a $z$-link \cite{ki}.
The ground state is vortex-free and the corresponding Hamiltonian
$H_0({\bf p})$ is given by
\begin{equation}
H_0({\bf p})=\left(\begin{array}{cc}
                  0&if({\bf p})\\
                  -if^*({\bf p})&0\\
                  \end{array}\right),
\end{equation}
with $f({\bf p})=2(J_xe^{i\bf p\cdot n_1}+J_ye^{i\bf p\cdot
n_2}+J_z)$. Here we still follow Kitaev and choose the basis of
the translation group ${\bf n}_{1,2}=(\pm 1/2,\sqrt{3}/2)$. In the
next section, we will see that deforming the angle between the
$x$-link and $y$-link to a rectangle will be much convenient.  The
eigenenergy may be obtained by diagonalizing the Hamiltonian,
which is $E_0({\bf p})=\pm |f({\bf p})|$. The phase diagram of the
model has been figured out in Fig. 1. Kitaev calls the gapped
phase as A-phase and the gapless phase B phase. The A phase is
topologically trivial and gapped. It is the strong-coupling limit
of SU(2) like the antiferromagnetic Heisenberg model \cite{aff}
and can be explained as the strong paring phase in the $p$ wave
sence \cite{rg}. After perturbed by an external magentic field,
the B phase is gapped and  has a non-zero spectral Chern number
and then is topologically non-trivial\ cite{ki}. Without lose of
generality, we consider $J_x=J_y=J_z=J$.
 The effective Hamiltonian is then given by $ H_0({\bf
p},\Delta)=(-f_2({\bf p}))\sigma^x+(-f_1({\bf
p}))\sigma^y+\Delta({\bf p}) \sigma^z$ with $\Delta(-{\bf
p})=-\Delta({\bf p})$,
$f_1=2J+4J\cos\frac{1}2p_x\cos\frac{\sqrt{3}}2p_y$ and
$f_2=4J\cos\frac{1}2p_x\sin\frac{\sqrt{3}}2p_y$ ($f=f_1+if_2$).
Assume $\psi^\pm({\bf p})$ to be the solutions of Schrodinger
equation $H_0({\bf p},\Delta)\psi^\pm({\bf p})=\pm E({\bf
p})\psi^\pm({\bf p})$ with $E({\bf p})=\sqrt{|f|^2+|\Delta|^2}$.
After normalization, we have ${\bf L}\cdot {\vec \sigma}\psi^\pm=
\pm\psi^\pm$ with
\begin{eqnarray}
{\bf L}=\frac{1}{\sqrt{3}J E({\bf p})}(-{f_2}({\bf p}),-{f_1}({\bf
p}),{\Delta}({\bf p})).
\end{eqnarray}
Explicitly, near ${\bf p}={\bf p}_*=-\frac{2}{3}{\bf
p}_1+\frac{2}3{\bf p}_2$mod$({\bf p}_1,{\bf p}_2)$ with ${\bf
p}_1$ and ${\bf p}_2$ the dual vectors of ${\bf n}_1$ and ${\bf
n}_2$, it is $ {\bf L}=\frac{1}{\hat E({\bf p})}(\delta p_y,\delta
p_x,\frac{\Delta({\bf p}_*)}{\sqrt{3}J})\equiv(\delta \hat
p_y,\delta \hat q_x,\hat\Delta)$ with $\hat E({\bf
p})=\sqrt{(\delta p_x)^2+(\delta p_y)^2+\Delta^2/3J^2}$. Near
${\bf p}=-{\bf p}_*$, it is $ {\bf L}=(\delta \hat p_y,-\delta
\hat p_x,-\hat\Delta). $ According to Kitaev, one can define a
spectrum Chern number by using the vector field ${\bf L}$. We will
be back to this issue when studying the index theorem.

\begin{figure}[htb]
\begin{center}
\includegraphics[width=4cm]{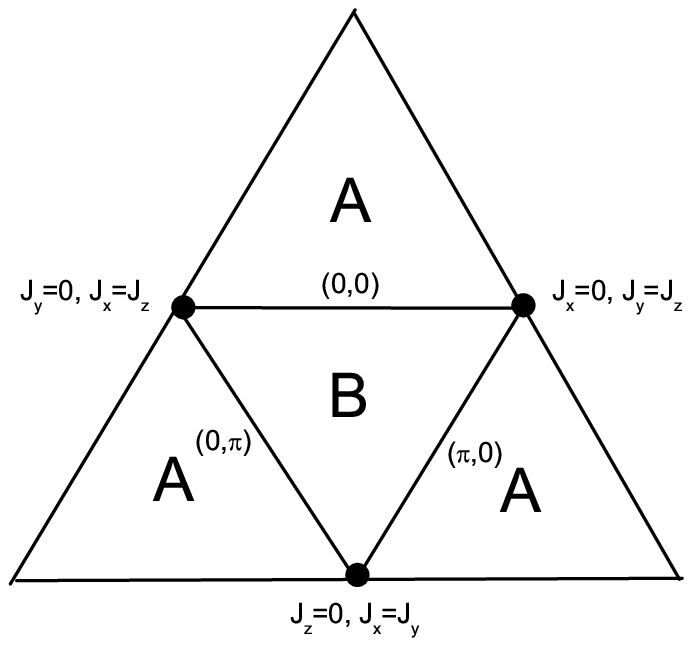}
\end{center}
 \caption{\label{fig:Fig. 1} Phase diagram in $( J_x,J_y,J_z)$-space. }
This is a (1,1,1)-cross section in all positive region.
\end{figure}

\section{Generalized exactly soluble model}

\begin{figure}[htb]
\begin{center}
\includegraphics[width=4cm]{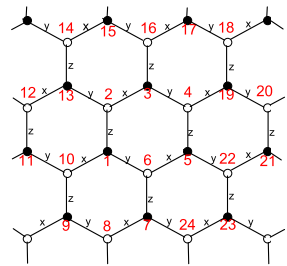}
\end{center}
 \caption{\label{fig:Fig. 2} (Color online): The honeycomb lattices and links.}
\end{figure}

We now consider the Hamiltonian which is generalization of the
Kitaev model in honeyconmb lattice to the following one
\begin{eqnarray}
H&=&-J_x\sum_{x-links}\sigma_i^x
\sigma_j^x-J_y\sum_{y-links}\sigma_i^y
\sigma_j^y-J_z\sum_{z-links}\sigma_i^z \sigma_j^z\nonumber\\
&-&\kappa_x\sum_b
\sigma^z_b\sigma^y_{b+e_z}\sigma^x_{b+e_z+e_x}\nonumber\\&
-&\kappa_x\sum_w
\sigma^x_{w}\sigma^y_{w+e_x}\sigma^z_{w+e_x+e_z}\nonumber\\&-&\kappa_y\sum_b
\sigma^z_{b}\sigma^x_{b+e_z}\sigma^y_{b+e_z+e_y}\nonumber\\
&-&\kappa_y\sum_w
\sigma^y_{w}\sigma^x_{w+e_y}\sigma^z_{w+e_y+e_z}\nonumber\\
&-&\lambda_x\sum_{b}\sigma_{b}^z\sigma^y_{b+e_z}
\sigma_{b+e_z+e_x}^y\sigma^z_{b+e_z+e_x+e_z}\nonumber\\&-&\lambda_y\sum_{b}
\sigma_{b}^z\sigma^x_{b+e_z}
\sigma_{b+e_z+e_y}^x\sigma^z_{b+e_z+e_y+e_z}\label{spinH}\\
&-&B_b\sum_{b}\sigma^z_b\sigma^y_{b+e_z}\sigma^z_{b+e_z+e_x}\sigma^y_{b+e_z+e_x-e_y}\nonumber\\
&-&B_w\sum_w\sigma_w^x\sigma_{w+e_x}^z\sigma_{w+e_x-e_y}^x\sigma_{w+e_x-e_y-e_z}^z
\nonumber\\
&-&B_w\sum_b\sigma_{b-e_y}^y\sigma_{b}^x\sigma_{b+e_z}^y\sigma_{b+e_z-e_x}^x
\nonumber\\
&-&
B_b\sum_b\sigma^z_{b-e_y-e_z}\sigma^x_{b-e_y}\sigma^x_b\sigma^y_{b+e_z}
\sigma^y_{b+e_z+e_x}\sigma^z_{b+e_z+e_x+e_z}\nonumber
\end{eqnarray}
where $'w'$ and $'b'$ labels the white and black sites in lattice
and $e_x,e_y,e_z$ are the positive unit vectors, which are defined
as, e.g., $e_{12}=e_z,e_{23}=e_x,e_{61}=e_y$ (See Fig.
\ref{fig:Fig. 2}). $J_{x,y,z}$, $\kappa_{x,y}$, $\lambda_{x,y}$
and $B_{b,w}$ are real parameters. This is a generalization of
Kitaev model  with the three-spin,four-spin and six-spin terms. It
is easy to check this generalized Hamiltonian still has a $Z_2$
gauge symmetry acting by a group element, e.g.,
$W_P=\sigma^x_1\sigma^y_2\sigma^z_3\sigma^x_4\sigma^y_5\sigma^z_6$
with $[H,W_P]=0$. In fact, one can add more $Z_2$ gauge invariant
multi-spin terms, e.g.,
\begin{eqnarray}
&&\sigma_{9}^z\sigma_{10}^y\sigma^y_1\sigma^y_2\sigma^x_3,\nonumber\\
&&\sigma_{9}^z\sigma_{10}^y\sigma^y_1\sigma^y_2\sigma^z_3\sigma^y_4,\nonumber\\
&&\sigma_{9}^z\sigma_{10}^y\sigma^y_1\sigma^y_2\sigma^y_3\sigma^z_{16},\nonumber
\end{eqnarray}
and so on. The site indices are shown in Fig. \ref{fig:Fig. 2}.
For our purpose, however, we restrict on (\ref{spinH}).

\begin{figure}[htb]
\begin{center}
\includegraphics[width=4cm]{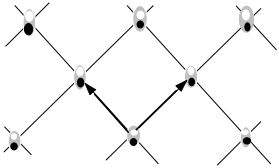}
\end{center}
 \caption{\label{fig:Fig. 3}
The effective square lattice. }
\end{figure}

We now use the Majorana fermion representation for this spin model
and then the Hamiltonian reads
\begin{eqnarray}
H &=&i\sum_a \sum_{a-links} J_au^a_{ij} c_ic_j
-i\sum_b K^x_{b,b+e_z}c_bc_{b+e_z+e_x}\nonumber\\
&-&i\sum_wK^x_{w+e_x-e_x,w-e_z-e_x}c_{w+e_x-e_x}c_{w-e_z-e_x}\nonumber\\
&-&i\sum_{b}\Lambda^x_{b,b+2e_z+e_x}c_{b}c_{b+2e_z+e_x}
\nonumber\\&-&i\sum_{w}\Lambda^x_{w,w-2e_z-e_x}c_{w}c_{w-2e_z-e_x}
\nonumber\\&+&y{\rm -partners}\nonumber\\
&+&i\sum_b \beta_{b,b+e_z+e_x-e_y} c_b
c_{b+e_z+e_x-e_y}\nonumber\\
&+&i\sum_w \alpha_{w,w+e_x-e_y-e_z} c_w
c_{w+e_x-e_y-e_z}\nonumber\\
&+&i\sum_w
\alpha_{w,b+e_y+e_x}c_{w}c_{b+e_y+e_x}\nonumber\\
&+&i\sum_b \beta_{b,b+e_z+e_y+e_x+e_z}c_{b}c_{b+e_z+e_x+e_y+e_z}
\end{eqnarray}
where $K^x_{b,b+e_z}=\kappa_x
u^z_{b,b+e_z}u^x_{b+e_z+e_x,b+e_z},~\Lambda^x_{b,b+2e_z+e_x}=\lambda_x
u^z_{b,b+e_z}u^x_{b+e_z,b+e_z+e_x}u^z_{b+e_z+e_x,b+e_z+e_x+e_z}$
etc and $u^a_{ij}=ib_i^ab_j^a$ on $a$-links.
$\beta_{b,b+e_z+e_x-e_y}=
B_bu^z_{b,b+e_z}u_{b+e_z,b+e_z+e_x}^xu_{b+e_z+e_x,b+e_z
+e_x-e_y}^y$ and $\alpha_{w,w+e_x-e_y-e_z}=B_w u^x_{w,w+e_x}
u_{w+e_x,w+e_x-e_y}^y$ $ u_{w+e_x-e_y,w+e_x-e_y-e_z}^z$ etc. It
can be shown that the Hamiltonian commutes with $u^a_{ij}$ and
thus the eigenvalues of $u_{ij}=\pm 1$. Since the four spin and
six-spin terms we introduced are related to the hopping between
the 'b' and 'w' sites, Lieb's theorem \cite{lieb} is still
applied. Following Kitaev, we take $u_{bw}=-u_{wb}=1$ and the
vortex-free Hamiltonian is given by
\begin{eqnarray}
H_0 &=&i\tilde
J_x\sum_s(c_{sb}c_{s-e_x,w}-c_{s,w}c_{s-e_x,b})\nonumber\\&+&i\tilde
\lambda_x\sum_s(c_{s,b}c_{s-e_x,w}+c_{s,w}c_{s-e_x,b}) \nonumber\\
&+&i\frac{\kappa^x}2\sum_s
(c_{s,b}c_{s+e_x,b}+c_{s,w}c_{s-e_x,w})\nonumber\\
&+&y~{\rm partners}+ iJ_z\sum_s
c_{sb}c_{sw}\nonumber\\
&+&iB_b\sum_s c_{s,b}
c_{s+e_x-e_y,w}+iB_w\sum_s c_{s,w}c_{s+e_x-e_y,b} \nonumber\\
&+&iB_w\sum_s c_{s,w} c_{s+e_y+e_x,b}+iB_b\sum_s
c_{s,b}c_{s+e_x+e_y,w}\nonumber
\\
 &=&i\tilde
J_x\sum_s(c_{sb}c_{s-e_x,w}-c_{s,w}c_{s-e_x,b})\nonumber\\&+&i\tilde
\lambda_x\sum_s(c_{s,b}c_{s-e_x,w}+c_{s,w}c_{s-e_x,b}) \nonumber\\
&+&i\frac{\kappa^x}2\sum_s
(c_{s,b}c_{s+e_x,b}+c_{s,w}c_{s-e_x,w})\nonumber\\
&+&y~{\rm partners}+ iJ_z\sum_s
c_{sb}c_{sw}\nonumber\\
&+&iB^-\sum_s (c_{s,b}
c_{s+e_x-e_y,w}-c_{s,w}c_{s+e_x-e_y,b}) \\
&+&iB^+\sum_s (c_{s,b}
c_{s+e_x-e_y,w}+c_{s,w}c_{s+e_x-e_y,b}) \nonumber\\
&+&iB^-\sum_s (c_{s,b}
c_{s+e_x+e_y,w}-c_{s,w}c_{s+e_x+e_y,b}) \nonumber\\
&+&iB^+\sum_s (c_{s,b}
c_{s+e_x+e_y,w}+c_{s,w}c_{s+e_x+e_y,b})\nonumber
\end{eqnarray}
where $s$ represents the position of a $z$-link, $\tilde
\lambda_\alpha=\frac{J_\alpha+\lambda_\alpha}2$ and $\tilde
J_\alpha=\frac{J_\alpha-\lambda_\alpha}2$. $B^\pm=\frac{B_b\pm
B_w}2$.

To simplify the pairing, one takes $B^+=0$ and denotes $B\equiv
B^-$. Defining a fermion on $z$-links by \cite{cn,yw}
\begin{eqnarray}
d_s=(c_{s,b}+ic_{s,w})/2, ~~~d^\dag_s=(c_{s,b}-ic_{s,w})/2,
\end{eqnarray}
the vortex-free Hamiltonian becomes an effective model of spinless
fermions on a square lattice (Fig.\ref{fig:Fig. 3})
\begin{eqnarray}
H_0&=&J_z\sum_s(d_s^\dag d_s-1/2)\nonumber\\
&+& B\sum_s(d_s^\dag d_{s\pm e_x\pm e_y}-d_sd^\dag_{s\pm e_x\pm
e_y}) \nonumber\\&+&\tilde
J_x(d^\dag_sd_{s+e_x}-d_sd_{s+e_x}^\dag)\nonumber\\&+& \tilde
\lambda_x\sum_s(d^\dag_{s+e_x}d^\dag_s-d_{s+e_x}d_s)\nonumber\\&+&
i\kappa_x\sum_s(d_sd_{s+e_x}+d_s^\dag d_{s+e_x}^\dag)+y~{\rm
partners}.
\end{eqnarray}
Or it is
\begin{eqnarray}
&& H_0=\sum_{\langle ij\rangle}(-td_i^\dag d_j+\Delta_{ij}d^\dag_i
d^\dag_{ij}+{\rm h.c.})-\mu\sum_i(d_i^\dag d_i-1/2)\nonumber\\
&&-t'\sum_{\langle\langle ij\rangle\rangle}(d_i^\dag d_j+{\rm
h.c.})+\delta\sum_{i,\pm}(d_i^\dag d_{i\pm e_x}-d_i^\dag d_{i\pm
e_y}), \label{antip}
\end{eqnarray}
where $t=-\frac{\tilde J_x+\tilde J_y}2$, $t'=-B$, $\mu=-J_z$ and
$\delta=\frac{\tilde J_x-\tilde J_y}2$. The paring parameters are
defined by $\Delta_{i,i\pm e_{x,y}}=\tilde
\lambda_{x,y}+i\kappa_{x,y}$. The last equality in
eq.(\ref{antip}) is the toy model Hamiltonian describing
Pfaffian/anti-Pfaffian states \cite{pfapf1}. Note that the pairing
free Hamiltonian is particle-hole symmetry if
$\Delta_{ij}=\mu=t'=0$. The $t'$-term breaks the particle-hole
symmetry. The $\delta$-term breaks the $\pi/2$ rotational
symmetry.  After Fourier transformation $d_s=\frac{1}{\sqrt
L_xL_y}\sum_{\bf p}e^{i{\bf p \cdot s}} d_{\bf p}$, we have
\begin{eqnarray}
H_0&=&\sum_{\bf p} \xi_{\bf p}d^\dag_{\bf p}d_{\bf
p}+\frac{\Delta^1_{\bf p}}2(d_{\bf p}^\dag d^\dag_{-{\bf p}}+d_{\bf p} d_{-{\bf p}})\nonumber\\
&+&i\frac{\Delta^2_{\bf p}}2(d_{\bf p}^\dag d^\dag_{-{\bf
p}}-d_{\bf p} d_{-{\bf p}})
\end{eqnarray}
where the dispersion relation is
\begin{eqnarray}
\xi_{\bf p}=J_z-\tilde J_x\cos p_x-\tilde J_y\cos p_y+2B\cos
p_x\cos p_y
\end{eqnarray}
 and the
pairing functions are
\begin{eqnarray}
&&\Delta^1_{\bf p}=\Delta_{1x}\sin p_x+\Delta_{1y}\sin
p_y,\nonumber\\
&&\Delta^2_{\bf p}=\Delta_{2x}\sin p_x+\Delta_{2y}\sin p_y
\end{eqnarray}
 with
$\Delta_{1,x(y)}=\kappa_{x(y)}$ and
$\Delta_{2,x(y)}=\tilde\lambda_{x(y)}$. After Bogoliubov
transformation,
\begin{eqnarray}
&&\alpha_{\bf p}=u_{\bf p}d_{\bf p}-v_{\bf p}d^\dag_{-\bf
p},\nonumber\\ &&\alpha^\dag_{\bf p}=u^*_{\bf p}d^\dag_{\bf
p}-v^*_{\bf p}d_{-\bf p}
\end{eqnarray}
 the Hamiltonian can be diagonalized
\begin{eqnarray}
H_0=\sum_{\bf p} E_{\bf p}\alpha^\dag_{\bf p}\alpha_{\bf p}+{\rm
const.}
\end{eqnarray}
 and the Bogoliubov
quasiparticles have the dispersion
\begin{eqnarray}
E_{\bf p}=\sqrt{\xi_{\bf p}^2+(\Delta^1_{\bf p})^2+(\Delta^2_{\bf
p})^2}.
\end{eqnarray}
The Bogoliubov-de Gennes equations are given by
\begin{eqnarray}
E_{\bf p}u_{\bf p}=\xi_{\bf p}u_{\bf p}-\Delta^*_{\bf p}v_{\bf
p},~~E_{\bf p}v_{\bf p}=-\xi_{\bf p}v_{\bf p}-\Delta_{\bf p}u_{\bf
p}
\end{eqnarray}
with
\begin{eqnarray}
&&v_{\bf p}/u_{\bf p}=-(E_{\bf p}-\xi_{\bf p})/\Delta^*_{\bf
p},\nonumber\\
&&|u_{\bf p}|^2=\frac{1}2(1+\frac{\xi_{\bf p}}{E_{\bf
p}}),\nonumber\\
&&|v_{\bf p}|^2=\frac{1}2(1-\frac{\xi_{\bf p}}{E_{\bf p}}).
\end{eqnarray}

\section{Phase diagram}

We now study the phase diagram in parameter space. The phase
diagram when $t'=0$ has been discussed in our previous work
\cite{yw}, which has the same shape as that in original Kitaev
model (with $(J_x,J_y,J_z)$ in Fig.1 substituted by $(\tilde
J_x,\tilde J_y,J_z)$) but the structures of the B phase are more
fruitful. After including the $t'$-term, the phase boundary is
still in $(p_x,p_y)=(0,0),(0,\pm\pi),(\pm \pi,0),(\pm \pi,\pm
\pi)$ as we know before. For the present model, it is $\tilde
J_z\pm \tilde J_x \pm \tilde J_y=0$ with $\tilde J_z=J_z+ 2B$ for
$(0,0),(\pm \pi,\pm \pi)$ and $J_z-2B$ for $(0,\pm\pi),(\pm
\pi,0)$. In $(\tilde J_x,\tilde J_y,\tilde J_z)$ space, the phase
diagram are of the same shape as that of original Kitaev model
(Fig. \ref{fig:Fig. 1}, $(J_x,J_y,J_z)$ is replaced by $(\tilde
J_x,\tilde J_y,\tilde J_z)$ ). The A phase is a strong pairing
phase. The nature of the B phase is much more intriguing. Inside
the B phase, $\xi_p$, $\Delta_{1,{\bf p}}$ and $\Delta_{2,{\bf
p}}$ can be zero individually. The gapless condition ($E_p=0$)
requires all three to be zero at a common ${\bf p}^*$. This can
only be achieved if (i) one of the $\Delta_{a,\bf p}=0$ or (ii)
$\Delta_{1,\bf p}\propto \Delta_{2,\bf p}$. If either (i) or (ii)
is true, $\xi_{\bf p}$ and $\Delta_{\bf p}$ can vanish
simultaneously, i.e. $E_p=0$ at ${\bf p}^*$, and the paired state
is gapless. Otherwise, the B phase is gapped. Note that contrary
to conventional wisdom, T-symmetry breaking alone does not
guarantee a gap opening in the B phase. The symmetry reason behind
the gapless condition of the B phase becomes clear in the
continuum limit where $E_p=0$ implies that the vortex-free
Hamiltonian must be invariant, up to a constant, under the
transformation $p_x\leftrightarrow \eta p_y$ and $\tilde
J_x\leftrightarrow \eta^{-2} \tilde J_y$ with
$\eta=\frac{\Delta_{a,y}}{\Delta_{a,x}}$ with $a=1$ or $2$ and for
nonzero $\Delta$. We refer to this as a {\em generalized inversion
(G-inversion) symmetry} since it reduces to the usual mirror
reflection when $\eta=1$. This (projective) symmetry protects the
gapless nature of fermionic excitations and may be associated with
the underlying quantum order \cite{wz}. Kitaev's original model
has $\Delta_{1,i}=0$, and is thus G-inversion invariant and
gapless. The magnetic field perturbation \cite{ki} breaks this
G-inversion symmetry and the fermionic excitation becomes gapped.

The continuum limit takes place near the critical line (0,0). In
this case,
\begin{eqnarray}
\tilde J_z-\tilde J_x-\tilde J_y-2t'=-\mu_{eff}-2t'=0.
\end{eqnarray}
with $\mu_{eff}=J_z-\tilde J_x-\tilde J_y$. Slight inside of the B
phase,
\begin{eqnarray}
\mu_{eff}\gtrsim -2t'.
\end{eqnarray}
If $t'<0$, $\mu_{eff}$ is positive and the system is in the
Pfaffian phase. If $t'>0$,
\begin{eqnarray}
\mu_{eff}\lesssim 2|t'|
\end{eqnarray}
 may be either positive or negative.
$\mu_{eff}>0$  means that the electron Fermi surface is closed
($\Delta_{ij}=0$) while the hole Fermi surface is opened. This is
the $d$-particle paired phase. On the other hand, $\mu_{eff}<0$
means that the electron Fermi surface is opened while the hole
Fermi surface is closed. This is the $d$-hole paired phase. Thus,
there is a Pfaffian/anti-Pfaffian transition as $\mu_{eff}$ is
across zero.

There are other two critical lines $(0,\pi)$ and $(\pi,0)$ near
which there are also gapless excitations. The critical condition
is given by
\begin{eqnarray}
&&J_z\mp \tilde J_x\pm \tilde J_y-2B=-\mu\mp\delta+2t'\nonumber\\
&&=2t'-\mu_{eff}=0 .
\end{eqnarray}
Inside of the B phase but near the critical lines,
\begin{eqnarray}
\mu_{eff}-2t'\gtrsim 0.
\end{eqnarray}
To satisfy this condition, $\mu_{eff}$ has to be the same sign as
that of $t'$.  Therefore, if $t'>0$, the system is in the Pfaffian
state and the system is in anti-Pfaffian state if $t'<0$.

We see that when $t'<0$, the sign of the effective chemical near
the critical line $(0,0)$ has the opposite dependence on the sign
of $t'$  to the effective chemical near other two critical lines.
For $t'>0$, the sign of the effective chemical near the critical
line $(0,0)$ may change from the opposite to the same as that near
other two critical lines. Hence, if $t'\ne 0$, there must be a
Pfaffian/anti-Pfaffian transition inside of the gapped B phase.

At $t'=0$, the Pfaffian and anti-Pfaffian states are degenerate.
As we have seen \cite{yw}, in the gapless B phase, there are two
gapless Majorana excitations at nodal points while in the gapped B
phase, the particle-hole symmetry is spontaneously broken, i.e.,
the ground state is either Pfaffian or anti-Pfaffian. All above
discussions are consistent with those in ref. \cite{pfapf1}.

\section{Continuous limit, Dirac equations and SO(3) gauge
theory}

 In fact, the
ground state wave function for a general $p$-wave paired state can
also be calculated in the continuous limit. The BCS wave function
is given by
\begin{eqnarray}
|\Omega\rangle=\prod_{\bf p}|u_{\bf
p}|^{1/2}\exp{(\frac{1}2\sum_{\bf p} g_{\bf p}d^\dag_{\bf
p}d^\dag_{\bf -p})}|0\rangle,
\end{eqnarray}
where $g_{\bf p}=v_{\bf p}/u_{\bf p}$. For even fermion number
$N$, the Pfaffian ground state wave function reads
\begin{eqnarray}
\Psi({\bf r}_1,...,{\bf r}_1)\propto \sum_P{\rm sgn}
P\prod_{i=1}^{N/2}g({\bf r}_{P_{2i-1}}-{\bf r}_{P_{2i}})
\end{eqnarray}
with $g_{\bf p}$ is the Fourier transform of $g({\bf r})$. For the
A phase, $g_{\bf p}\propto \Delta_{\bf p}$ and the analyticity of
$g_{\bf p}$ leads to $g({\bf r})\propto e^{-\mu r}$ as the same
calculation in a pure $p_x+ip_y$ strong pairing state. In the B
phase, if the G-chiral symmetry is broken, defining $
p_a'=\Delta_{ab}p_b$ with $a=1,2$ and $b=x,y$, $g_{\bf p}\propto
\frac{1}{p_x'+ip_y'}$ and then $g({\bf r})=\frac{1}{x_1'+ix_2'}$
with $x'_a=\Delta^{-1}_{ab}x_b$. This is a Pfaffian state with
$z'=x'_1+ix_2'$ and is corresponding to a weak paired gapped
fermion state \cite{rg}.

 In the long
wave length limit (small $p$ limit), we approximate $\xi_{\bf
p}\approx-m=\tilde J_z-\tilde J_x-\tilde J_y$ and define
$\psi(t,{\bf s})=(u(t,{\bf s}),v(t,{\bf s}))$ with $(u,v)$ the
Fourier transformation of $(u_{\bf p},v_{\bf p})$. The BdG
equations in the gapped B phase reads
\begin{eqnarray}
e^{\mu a}\gamma_a\partial_\mu \psi+im\psi=0
\end{eqnarray}
with the vielbein $e^{\mu t}=\delta_{\mu t}$ and
$e^{ij}=\Delta_{ij}$. $\gamma_0=\sigma^z$ and
$\gamma_i=\sigma^z\sigma^i$. Furthermore, it may be generalized to
a curve space with spin connection term added \cite{rg}. In
general, the equation of $u$ is not compatible to that of $v$ and
the fermions are Dirac fermions. However, it is easy to show that
for this $p$-wave paired state, $u$ and $v^*$ obey the same
equation, i.e., the anti-particle of the quasiparticle is itself.
Then the fermions are Majorana ones.

We now turn to local gauge transformation. As we have pointed out,
taking $u_{ij}=1$ means the whole SU(2) gauge symmetry is fixed.
If making an SU(2) gauge transformation, we will run out the
vortex-free state. Keeping in the vortex-free state, the gauge
transformation is confined in an SU(2)/$Z_2\sim$ SO(3) one.
Therefore, making an SO(3) transformation,
\begin{eqnarray}
\psi'({\bf r},t)=U({\bf r},t)\psi({\bf r},t),
\end{eqnarray}
Dirac equations read
\begin{eqnarray}
e^{\mu a}\gamma_aD_\mu \psi'+im\psi'=0,
\end{eqnarray}
where the covariant derivative $D_\mu=\partial_\mu+A_\mu$ with
$A_\mu=U^{-1}\partial_\mu U$ with $U\in$SO(3). This gauge
potential is a pure gauge with a vanishing strength $F=dA+A\wedge
A=0$.

We now discuss the nature of the gapless B phase in the general
model with G-inversion symmetry. In this case, $E_{\bf p}=0$ at
${\bf p}=  \pm {\bf p}^*$ which are the solutions of $\xi_{\bf
p}=0$ and, say, $\Delta_{\bf p}=\Delta_{1,{\bf p}}=0$. At ${\bf
p}^*$, the fermion dispersions are generally given by 2D Dirac
cones. However, by a continuous variation of the parameters, one
can realize a dimensional reduction near the phase boundary where
the effective theory is in fact a 1+1 dimensional conformal field
theory in the long wave length limit. Let us consider parameters
that are close to the critical line with $|\sin p_a^*|\ll |\cos
p_a^*|$ where $g_{\bf q}={\rm sgn}[q_x\Delta_{1x}\cos
p_x^*+q_y\Delta_{1y}\cos p_y^*]\equiv{\rm sgn}(q_x')$ with ${\bf
q}={\bf p}-{\bf p}^*$. Doing the Fourier transform, we find
\begin{eqnarray}
g({\bf r})&=&\int dq'_xdq'_y e^{iq_x'x'+iq_y'y}{\rm
sgn}(q_x')\nonumber\\
&=&\delta(y') \int dq_x' \frac{q_x'}{|q_x'|}\sin q_x'x'\sim
\frac{\delta(y')}{ x'}.
\end{eqnarray}
The $\delta(y)$-function indicates that the pairing in the gapless
B phase has a one-dimensional character and the ground state is a
one-dimensional Moore-Read Pfaffian. The BdG equations reduces to
\begin{eqnarray}
i\partial_t u=-i\Delta_{1x}(1+i\eta)\partial_{x'}v,~~i\partial_t
v=i\Delta_{1x}(1-i\eta)\partial_{x'}u,
\end{eqnarray}
with $\eta=\frac{\Delta_{1y}}{\Delta_{1x}}$. Thus, the gapless
Bogoliubov quasiparticles are one-dimensional Majorana fermions.
The long wave length effective theory for the gapless B phase near
the phase boundary is therefore the massless Majorana fermion
theory in 1+1-dimensional space-time, i.e. a $c=1/2$ conformal
field theory or equivalently a two-dimensional Ising model.

\section{Topological invariants and Index theorem}

\subsection{Spectral Chern Number and $\eta$-invariant}

 We note that there is no spontaneous breaking of a
continuous symmetry in the phase transition from A to B phases.
Kitaev has shown that the A phase in his model is topological
trivial and has zero spectral Chern number while the gapped B
phase has this Chern number $\pm 1$ \cite{ki}. This fact was also
already realized by Read and Green in the $p_x+ip_y$ paired state.
Here we follow Read and Green \cite{rg} to study this topological
invariant for a general $p$-wave state.

In continuous limit, ${\bf p}=(p_x,p_y)$ is in an Euclidean space
$R^2$. However, there is a constraint $|u_p|^2+|v_p|^2=1$, which
parameterizes a sphere $S^2$. As $|{\bf p}|\to\infty$, $\xi_{\bf
p}\to E_{\bf p}$. Then, $v_{\bf p}\to 0$ as $|{\bf p}|\to\infty$.
Therefore, we can compact $R^2$ as an $S^2$ by adding $\infty$ in
which $v_{\bf p}\to 0$ to $R^2$. The sphere $|u_{\bf p}|^2+|v_{\bf
p}|^2=1$ can also be parameterized by ${\bf
n_p}=(\Delta^{(1)}_{\bf p},-\Delta^{(2)}_{\bf p},\xi_{\bf
p})/E_{\bf p}$ because $|{\bf n_p}|=1$. $(u_{\bf p},v_{\bf p})$
describes a mapping from $S^2~ ({\bf p}\in R^2)$ to $S^2$ (spinor
$|{\bf n_p}|=1$). The winding number of the mapping is a
topological invariant.The north pole is $u_{\bf p}=1,v_{\bf p}=0$
at $|{\bf p}|=\infty$ and south pole $u_{\bf p}=0,v_{\bf p}=1$ at
${\bf p=0}$. For ${\bf n_p}$ parametrization, ${\bf
n_0}=(0,0,\frac{\xi_{\bf p}}{E_{\bf }})$ at $|{\bf p}|=\infty$ and
$(0,0,\frac{\xi_{\bf p}}{E_{\bf }})$ at ${\bf p}=0$, either the
north pole or south pole.

For strong pairing phase, we know that $u_p\to 1$ and $v_p\to 0$
as ${\bf p}\to 0$ (or equivalently, $\xi_{\bf p}>0$). This means
that for arbitrary ${\bf p}$, $(u_{\bf p},v_{\bf p})$ maps
$p$-sphere to the up-hemishpere and then winding number is zero.
That is, the topological number is zero in the strong pairing
phase.

For the weak paring phase, $u_{\bf p}\to 0$ and $v_{\bf p}\to 1$
as ${\bf p}\to 0$. This means that the winding number is not zero
(at least wrapping once). For our case, it may be directly
calculated  and the winding number is given by
\begin{eqnarray}
\nu&=&\frac{1}{4\pi}\int dp_xdp_y \partial_{p_x}{\bf n_p}\times
\partial_{p_y}{\bf n_p}\cdot {\bf n_p}=
 1
\end{eqnarray}
Defining $P({\bf p})=\frac{1}2(1+{\bf n_p}\cdot {\vec \sigma})$,
which is the Fourier component of the project operator to the
negative spectral space of the Hamiltonian, this winding number is
identified as the spectral Chern number defined by Kitaev
\cite{ki}
\begin{eqnarray}
\nu=\frac{1}{2\pi i} \int {\rm
Tr}(P_-(\partial_{p_x}P_-\partial_{p_y}P_-\partial_{p_y}P_-\partial_{p_x}P_-)]dp_xdp_y
.\end{eqnarray}

There is a topological invariant called Atiyah-Padoti-Singer
eta-invraiant \cite{aps} which reflects the asymmetry of the
spectrum of the Dirac operator
\begin{eqnarray}
&&\bar\eta(S^2)=\frac{1}2\lim_{s\to 0}\int d\lambda
\rho(\lambda){\rm sgn}(\lambda)|\lambda|^{-s},
\end{eqnarray}
where $\rho(\lambda)$ is the spectral density. Transforming the
variable from $\lambda$ to ${\bf p}$, the measure of the
integration from $\Delta_{\bf p}$ to ${\bf q}$ includes a Jacobian
determinant \cite{nie},
\begin{eqnarray}
&&\int d\lambda...=\frac{1}{4\pi}\int d \Delta^{(1)}_{\bf
p}d\Delta^{(1)}_{\bf p} \frac{1}{ E_{\bf
p}}...\nonumber\\&&=\frac{1}{4\pi}\int d^2p J({\bf \Delta_p,p})
\frac{1}{ E_{\bf p}}...\nonumber\\&&=\frac{1}{16\pi i} \int d^2p
{\rm Tr}[J( {\bf n_p}\cdot {\vec \sigma},{\bf q})...]\nonumber
\end{eqnarray}
Note that $-\frac{1}2{\bf n_p}\cdot {\vec \sigma}$ is the
signature matrix sgn$(H({\bf p}))$of the Hamiltonian, one has
\begin{eqnarray}
\bar\eta(S^2)&=&\frac{1}2\lim_{s\to 0}\int d\lambda
\rho(\lambda){\rm sgn}(\lambda)|\lambda|^{-s}\nonumber\\
&=&\int_{S^2}d{\bf p}{\rm Tr[\frac{1}2sgn}(H({\bf
p}))\rho(\frac{1}2{\rm sgn}(H({\bf p})))],
\end{eqnarray}
where
\begin{eqnarray}
\rho(\frac{1}2{\rm sgn}(H({\bf
q})))&=&\frac{i}{16\pi}\biggl(\frac{\partial {\rm sgn}(H({\bf
q}))}{dq_x}\frac{\partial {\rm sgn}(H({\bf
q}))}{dq_y}\nonumber\\&-&\frac{\partial {\rm sgn}(H({\bf
q}))}{dq_y}\frac{\partial {\rm sgn}(H({\bf q}))}{dq_x}\biggr).
\end{eqnarray}
Defining the spectral projector $I-P=P_-({\bf q})=\frac{1}2(1-{\rm
sgn} H({\bf q}))$, the eta-invariant is exactly equal to one half
of the spectral Chern number defined by Kitaev \cite{ki}
\begin{eqnarray}
&&\bar\eta(S^2)=\frac{1}{4\pi i}\int {\rm Tr}[P_-({\bf q})
dP_-({\bf q})\wedge d P_-({\bf q})].\label{index}
\end{eqnarray}
Kitaev identifies one half of $\nu$ as the chiral central charge
$c_-$. Our result shows that this chiral central charge is just
the eta-invariant. Physically, it is easy to be understood because
both $c_-$ and $\bar\eta$ reflect the anomaly of the spectrum of
the system.

\subsection{Index Theorem}

In the continuous limit, if the space is compacted as $S^2$, the
2+1 space-time is a ball $X$ with a boundary $B=S^2_+(\tau=1)\cup
Y\cup S^2_-(\tau=0)$ where $S^2_\pm$ are the top and bottom halves
of a sphere and $Y$ is a cylinder.  Now, we can apply the index
theorem (\ref{oi}) to this spectral problem of the Dirac operator
$D_\mu$ in $X$ with boundary $B$ \cite{daizh}. The general form of
thye index theorem in an odd-dimensional manifold is briefly
reviewed in Appendix A. In 2+1-dimensions, the index theorem for
the Toeplitz operator reads
\begin{eqnarray}
{\rm Ind}~T_g=\frac{1}{24\pi^2}\int_{X}{\rm
Tr}[(gdg^{-1})^3]-\bar\eta(B,g)+\tau_\mu(B,P,g) \label{3oi}
\end{eqnarray}
where the first term is equal to $\Gamma/2\pi i$ with $\Gamma$ the
WZ term. The Maslov triple index $\tau_\mu(B,P,g)$ is an integer
\cite{kl}.  We do not have a physical explanation of
$\tau_\mu(B,P,g)$ yet and it possibly relates to the central
charge of the theory \cite{turaev}. $\bar\eta$ is the reduced
eta-invariant. The first term in (\ref{3oi}) determines the bulk
state topological properties and the latter two terms reflect the
boundary topological properties. That the index ${\rm
Ind}~T_g-\tau_\mu$ is an integer determines the bulk-boundary
correspondence.

 The WZ term $\Gamma$ is defined for the
fundamental representation of SU(2) but $g$ is restricted to a
subgroup SU(2)/$Z_2\sim$ SO(3). The reduced eta-invariant is given
by
\begin{eqnarray}
&&\bar\eta(B,g)=\bar\eta(S^2_+,g_{\tau=1})
+\bar\eta(S^2_-,g_{\tau=0})\nonumber\\
&&=[\bar\eta(S^2_+,g_{\tau=1})-\bar\eta(S^2_+,g_{\tau=0})]\nonumber\\
&&~~~+[\bar\eta(S^2_+,g_{\tau=0})+\bar\eta(S^2_-,g_{\tau=0})]\nonumber
\\&&\equiv\Delta\bar\eta(S^2_+)+\bar\eta(S^2,g_{\tau=0}),
\end{eqnarray}
because $\bar\eta(Y)=0$ for $Y$ may contract to a cycle $S^1$. In
general, $\bar\eta(S^2_\pm,g_{1,0})=$$\frac{1}2[{\rm
dim(ker}D(S^2_\pm,g_{1,0}))+\eta(D(S^2_\pm,g_{1,0}))]. $ Because
of a non-zero gap, ${\rm dim(ker}D(S^2_\pm))=0$.
 Therefore,
$
\bar\eta(S^2,g)=\frac{1}2\Delta\eta(S^2_+)+\frac{1}2\eta(D(S^2,g_{\tau=0})).
$ The discrete eigenstates of the Dirac operator do not contribute
to $\bar\eta$ because there is no asymmetry of the spectrum for
these states. $\bar\eta(S^2,g_{\tau=0})=\bar\eta(S^2)$ is just the
eta-invariant calculated in the previous subsection.

 Now, the index theorem reads
\begin{eqnarray}
{\rm Ind}~T_g={\Gamma}/{2\pi
i}-\nu/2-\Delta\bar\eta(S^2_+)+\tau_\mu(B,P,g).
\end{eqnarray}
The integrity of $ {\rm Ind}~T_g-\tau_\mu(B,P,g)$ requires
\begin{eqnarray}
\Delta \Gamma/2\pi i\equiv{\Gamma}/{2\pi
i}-\Delta\bar\eta(S^2_+)=\nu/2 ~{\rm mod}(Z).
\end{eqnarray}
Dai and Zhang  have thought $\Delta\bar\eta(S^2_+)$ as an
intrinsic form of the WZ term \cite{daizh} and then $\Delta
\Gamma$ is in fact an ambiguity of the WZ term . It is seen
that,if $\nu$ is even, one requires $\Delta \Gamma=2\pi i\times$
integer. If $\nu$ is odd , $\Delta \Gamma/2\pi i$ is required to
be a half integer. In our model, it is known that for an SO(3)
group, $\Delta \Gamma=\pi i\times$integer, which is consistent
with $\nu=1$. In general, the index theorem (\ref{3oi}) gives a
constraint to the WZ term. An odd $\nu$ requires an even level $k$
WZ term in the effective action and the minimal one is $k=2$. This
is consistent with the non-abelian anyonic statistics of the
vortex excitations. For an even $\nu$, the minimal value of $k$ is
one, it is consistent with the abelian anyonic statistics.

 \vspace{0.1cm}

\section{Edge excitations and bulk-edge correspondence}

In the previous discussion, the two-dimensional space is taken to
be a tours without boundary. If we consider a two-dimensional
space with edge instead of the torus, Kitaev has shown that the
gapless chiral edge excitations coexist with a non-zero spectral
Chern number\cite{ki}. This is a general result if the bulk states
are gapped. In fact, it was generally known that the eta-invariant
of the Dirac operator can be related to the ground state fermion
charge \cite{jak,po,nie}. By using the continuous equation
$\partial_t\rho+\nabla\cdot {\bf j}=0$, the eat-invariant is
related to the edge current integrated along the one-dimensional
edge $S$, i.e.,\cite{nie}
\begin{eqnarray}
\eta=2\int_{S}d{\bf s}\cdot{\bf j}~,
\end{eqnarray}
where the factor '2' is different from '1' in (104) of Kitaev in
\cite{ki} because of $\frac{1}2$ factor in (\ref{ham}). Thus, a
non-zero eta-invariant corresponds to a non-vanishing net edge
current and then the gapless chiral edge excitations. Hence, the
index theorem already explained the bulk-edge correspondence.

\vspace{0.1cm}

\section{Conclusions }

We studied a generalized Kitaev model whose vortex-free sector can
be mapped to a $p$-wave paired state with the next nearest
neighbor hopping. The phase diagram is figured out. The property
of the gapped B phase was very interesting for a
Pfaffian/anti-Pfaffian phase transition was found in this phase.
According to the gauge invariance of the spin-1/2 theory in the
fermion representation, we found the low-lying effective theory of
the model is described by Majorana fermion coupled to a gauge
field. The existence of this non-dynamic gauge field enabled us to
understand the mathematic connotation behind these topological
orders. The edge conformal anomaly can be cancelled by the bulk WZ
term in terms of the recently proved index theorem on odd
manifold.

 \vspace{0.1cm}

\section{Acknowledgement}

The author would like to thank Z. Nussinov, N. Read, X. Wan, Z. H.
Wang, X.-G. Wen, T. Xiang, Y. S. Wu, K. Yang, J. W. Ye and M. Yu
for useful discussions. The author is grateful to Weiping Zhang
for him to explain how to understand the index theorem (\ref{oi}).
Especially, the author greatly appreciates Ziqiang Wang for our
successful cooperation in ref. \cite{yw}. Part of this work is the
generalization of our co-work. This work was supported in part by
the national natural science foundation of China, the national
program for basic research of MOST of China and a fund from CAS.

\appendix

\section{Index theorems: Mathematical preparation}

\label{app1} Index theorems relate the analytical index of a
differential operator to the topological index of a vector bundle
that the operator acts on. Atiyah and Singer (AS) proved a general
form of the index theorem for even dimensional compact manifold
\cite{as}. It said that the analytical index of an elliptical
differential operator on the vector bundle is a global topological
invariant which can be expressed by the integral of local
topological characters on the background manifold. A
generalization of the theorem to even manifolds with boundary,
so-called Atiyah-Patodi-Singer (APS) theorem \cite{aps}, e.g., for
a Dirac operator $D$ on a manifold $M$ with boundary $\partial M$,
reads
\begin{eqnarray}
{\rm Ind}D=\int_M\hat A(M)+\frac{1}2(h[\partial M]+\eta[\partial
M])+\omega(\partial M),\label{ei}
\end{eqnarray}
where $\hat A$ is the Hirzebruch $\hat A$-class of $M$, $h$ is the
dimensions of the zero modes of the boundary Dirac operator
$D_{\partial M}$ and $\eta$ is the APS eta-invariant defined by
$\eta(\partial M)=\lim_{s\to 0}\sum_{\lambda \ne 0}{\rm
sgn}(\lambda)|\lambda|^{-s}$ where $\lambda$ is the eigenvalue of
$D_{\partial M}$. $\omega(\partial M)$ is a Chern-Simons term
caused by the non-product boundary metric and physically giving
anomaly in quantum field theory \cite{zumino}. There are many
applications of the AS and APS theorem in physics, e.g, see
Refs.\cite{nie,po,jak}. Recently, index theorems were applied to
chiral $p$-wave superconductors \cite{lee} and graphene
\cite{stone1}.

There are partners of AS and APS theorems on odd-dimensional
manifolds $X$ ($d$=odd)\cite{odd1,odd2,daizh}. Again, we consider
Dirac operator $D$. ${\cal L}=\sum_\lambda {\cal L}_\lambda$ is
the spectrum space of $D$ where ${\cal L}_\lambda$ is the subspace
with eigenvalue $\lambda$. $P$ is a project operator defined by
$P{\cal L}={\cal L}_+=\sum_{\lambda\geq 0} {\cal L}_\lambda$. Let
${\cal L}$ trivially take its value on $C^N$, i.e., a state
$\psi\in {\cal L}$ is extended to an $N$ vector which transforms
under a group GL$(N,C)$. For Toeplitz operator $T_g=PgP$ on
odd-dimensional  manifolds $X$ with boundary $B$ in which the
group element $g$ is not an identity, an index theorem is given by
\cite{daizh}
\begin{eqnarray}
{\rm Ind}~T_g&=&-\frac{1}{(2\pi i)^{\frac{d+1}2}}\int _X\hat
A(R^X){\rm
Tr}[\exp(-R^{\cal L})]{\rm ch}(g)\nonumber\\
&-&\bar\eta[B,g]+\tau_\mu(B,P,g), \label{oi}
\end{eqnarray}
where $g\in$GL$(N,C)$ (or a subgroup like SU($N$));  $R^X$ and
$R^{\cal L}$ are the curvatures of the background manifold $X$ and
${\cal L}$; $\bar \eta[B,g]$ is a reduced eta-invariant and
$\tau_\mu$, which is an integer, is the Maslov triple index
\cite{kl}. We do not have a physical explanation of $\tau_\mu$ yet
and it possibly relates to the central charge of the theory
\cite{turaev}. ${\rm ch}(g)$ is odd Chern character defined by
\begin{eqnarray}
{\rm ch}(g)=\sum_{n=0}^{\frac{d-1}2}\frac{n!}{(2n+1)!}{\rm
Tr}[(g^{-1}dg)^{2n+1}].
\end{eqnarray}
The first term in (\ref{oi}) determines the bulk state topological
properties and the latter two terms reflect the boundary
topological properties. That the index ${\rm Ind}~T_g-\tau_\mu$ is
an integer determines the bulk-boundary correspondence.

\section{Spin and Majorana fermions }

In this appendix, we introduce the Majorana representation of
spin-1/2 operators. Consider the Pauli matrices
\begin{equation}
\sigma^x=\left(\begin{array}{cc}
0&1\\
1&0\\
\end{array}\right),\sigma^y=\left(\begin{array}{cc}
0&-i\\
i&0\\
\end{array}\right),\sigma^z=\left(\begin{array}{cc}
1&0\\
0&-1\\
\end{array}\right),
\end{equation}
with $\sigma^x\sigma^y=i\sigma^z$ and so on. The spin-1/2 matrices
are one-half of the Pauli matrices: ${\bf S}=\frac{\vec \sigma}2$.
Casimir of SU(2) group requires ${\bf S}\cdot {\bf S}=S(S+1)=3/4$
, i.e., ${\vec \sigma}\cdot {\vec \sigma}=3$. Using the
conventional fermion operators $c_\uparrow$ and $c_\downarrow$,
 the spin-1/2 operators can be expressed by
\begin{eqnarray}
 &&\hat\sigma^x=c^\dag_\uparrow c_\downarrow+c^\dag_\downarrow
 c_\uparrow\nonumber\\
 &&\hat\sigma^y=-i(c^\dag_\uparrow c_\downarrow-c^\dag_\downarrow
 c_\uparrow)\nonumber\\
 &&\hat\sigma^z=c^\dag_\uparrow c_\uparrow-c^\dag_\downarrow
 c_\downarrow.
 \end{eqnarray}
That is $\hat\sigma^a=c^\dag_s\sigma^a_{ss'}c_{s'}$. Since
$\{c,c^\dag\}=1,c^2={c^\dag}^2=0$, one has
 $\hat\sigma^x\hat\sigma^y=i\hat\sigma^z$. However, it is easy to
 see that
 $$(\hat\sigma^a)^2=n_{\uparrow}+n_{\downarrow}-2n_{\uparrow}n_{\downarrow}$$.
 To insure the Casimir operator constraint, one requires
$$n_{\uparrow}+n_{\downarrow}=1,~~n_{\uparrow}n_{\downarrow}=0$$
These are equivalent to an SU(2) constraint
\begin{eqnarray}
&&T^x=c^\dag_{\uparrow}c^\dag_{\downarrow}+c_{\downarrow}c_{\uparrow}=0,\nonumber\\
&&T^y=i(c^\dag_{\uparrow}c^\dag_{\downarrow}-c_{\downarrow}c_{\uparrow})=0,\nonumber\\&&
T^z=n_{\uparrow}+n_{\downarrow}-1=0
\end{eqnarray}
with $T^xT^y=iT^z$ and so on. It is well-known taht
$\{\hat\sigma^a/2,T^a/2\}$ form an$SO(4)\sim SU(2)\times
SU(2)/Z_2$ Lie algebra.

The fermion expression brings extra degrees of freedom and then
there is an SU(2) gauge invariant of the spin operators. Defining
\begin{equation}
(\chi_{\alpha\beta})=\left(\begin{array}{cc}
c_\uparrow&c_\downarrow\\
c^\dag_\downarrow&-c^\dag_\uparrow\\
\end{array}\right),
\end{equation}
the spin operator may be rewritten as
\begin{eqnarray}
{\hat \sigma^a}=\frac{1}2{\rm Tr}[\chi^\dag\chi(\sigma^a)^T]
\end{eqnarray}

Making a gauge transformation $\chi_{\alpha\beta}\to
g_{\alpha\gamma}\chi_{\gamma\beta}$ and then
$\chi^\dag_{\alpha\beta}\to
\chi^\dag_{\alpha\gamma}g^\dag_{\gamma\beta}$ for $g\in $SU(2) and
$g^\dag=g^{-1}$, one has $\hat \sigma^a$ is gauge invariant.

Majorana fermions are related to the conventional fermion through
\begin{eqnarray}
&&c_\uparrow=\frac{1}2(b_x-ib_y),~~c_\downarrow=\frac{1}2(b_z-ic)\nonumber\\
&&c^\dag_\uparrow=\frac{1}2(b_x+ib_y),~~c^\dag_\downarrow=\frac{1}2(b_z+ic)\nonumber
\end{eqnarray}
That is,
\begin{eqnarray}
&&b^x=c_\uparrow^\dag+c_\uparrow,~b^y=i(c_\uparrow^\dag-c_\uparrow),\nonumber\\
&&b^z=c_\downarrow^\dag+c_\downarrow,~c=i(c_\downarrow^\dag-c_\downarrow).\nonumber
\end{eqnarray}
It is easy to check that
\begin{eqnarray}
&&b_{a}^2=1,~~~c^2=1\nonumber\\
&&b_{a}b_{b}=-b_{b}b_{a}, cb_{a}=-b_{a}c
\end{eqnarray}
Therefore, using the Majorana fermions, one can express the spin
operators by
\begin{eqnarray}
&&\hat\sigma^x=\frac{i}2(b_xc-b_yb_z,\nonumber\\
&&\hat\sigma^y=\frac{i}2(b_yc-b_zb_x),\nonumber\\&&
\hat\sigma^z=\frac{i}2(b_zc-b_xb_y).
\end{eqnarray}
Correspondingly, the constraint reads
\begin{eqnarray}
&&T^x=\frac{i}2(b_xc+b_yb_z)=0,\nonumber\\&&T^y=\frac{i}2(b_yc+b_zb_x)=0,\nonumber\\
&&T^z=\frac{i}2(b_zc+b_xb_y)=0.
\end{eqnarray}
Compactly, it is $D=b_xb_yb_zc=1$. In an explicit gauge invariant
form, $D=-i\hat\sigma^x\hat\sigma^y\hat\sigma^z=1$. Using this
constraint, the spin operators are simplified as
\begin{eqnarray}
\hat\sigma^a=ib_ac.
\end{eqnarray}

\section{isolated vortex excitations}

The isolated vortex excitations above the Pfaffian ground state
has been discussed in \cite{yw}. For self-containing of the paper,
we repeat the paragraph which discuss the vortex excitations. The
$Z_2$ vortex excitation in the spin model which corresponds to
setting $W_P=-1$ for a given plaquette. Although the Hamiltonian
in the fermion representation is bilinear, it is difficult to
obtain analytical solutions of the wave function with vortex
excitations \cite{ki}. Our strategy is to evaluate the energy of
the Moore-Read \cite{mr} trial wave function with two well
separated half-vortices located at $w_1$ and $w_2$ shown in Fig.~4
. In the gapped B phase with $p_x+ip_y$-wave pairing,
\begin{eqnarray}
&&\Psi(z_1,...z_N;w_1,w_2)\propto {\rm
Pf}(g'(z_i,z_j;w_1,w_2)),\nonumber\\
&&g'(z_1,z_2;w_1,w_2)\propto\frac{(z_1-w_1)(z_2-w_2)+(w_1\leftrightarrow
w_2)}{z_1-z_2}.
\nonumber\\
&&|w_1,w_2\rangle\propto\exp\{\frac{1}2\sum_{{\bf r}_1,{\bf
r}_2}g'(z_1,z_2;w_1,w_2)d^\dag_{{\bf r}_1}d^\dag_{{\bf r}_2}\}
\end{eqnarray}
Performing a Fourier transformation, we have
\begin{eqnarray}
|w_1,w_2\rangle\propto\exp\{\frac{1}2\sum_{{\bf K},{\bf
k}}g'_k({\bf K})d_{\bf K+k}^\dag d_{\bf K-k}^\dag\},
\end{eqnarray}
where ${\bf k}={\bf k}_1-{\bf k}_2$ and ${\bf K}={\bf k}_1+{\bf
k}_2$ are the relative and the total momenta of the pairs and
$g'_k({\bf K})$ is the Fourier transform of $g'({\bf r}_1,{\bf
r}_2)$. One can show that, in a system with linear dimension $L$,
$g'_k(K=0)\sim \frac{1}k(1/6+i/8-(1+i)(w_1+w_2)/8L+w_1w_2/L^2)$.
The Hamiltonian in the presence of the two vortices shown in
Fig.~1 where the red $z$-links have $u_{bw}=-1$ and all others
$u_{bw}=1$, may be written as $H=H_0+\delta H$. Here $H_0$ is the
vortex-free Hamiltonian and $\delta H$ is the vortex part. The
latter is expressed as a sum of (twice) the pairing and chemical
potential terms in Eq.~(6) over the red $z$-links extending in the
$\xi$-direction (the line with $x=y$) between the vortices. By a
direct calculation, one can prove that $\delta H$ has the
following form
\begin{eqnarray}
&&\langle w_1,w_2|\delta H|w_1,w_2\rangle\\
&&\propto
\sum_{p_\xi,p'_\xi}\frac{i(e^{iw_1(p_\xi+p'_\xi)}-e^{iw_2(p_\xi+p'_\xi)})}
{p_\xi+p'_\xi}f(p_\xi,p_\xi')=0,\nonumber
\end{eqnarray}
where $f(p_\xi,p_\xi')$ is an analytical function of
$p_\xi+p'_\xi$. On the other hand, one can check that since
$[H_0,\sum_{\bf K,k}g'_k({\bf K}\ne 0)d_{\bf K+k}^\dag d_{\bf
K-k}^\dag]=0$, the ${\bf K}\ne0$ sector does not play a nontrivial
role in calculating $E_v$. The energy of such a vortex pair is
given by
\begin{eqnarray}
&&E_v=\langle w_1,w_2|H|w_1,w_2\rangle=\langle w_1,w_2|H_0|w_1,w_2
\rangle\nonumber\\
&&=\sum_{\bf k}E_{\bf k}|u_{\bf k}\delta g_{\bf k}|^2\langle
w_1w_2|d_{\bf k}d^\dag_{\bf k}|w_1w_2\rangle\nonumber\\
&&\sim A(w_1,w_2)\sum_{\bf k}(1-|g'_{\bf k}(K=0)|^2)E_{\bf k},
\end{eqnarray}
where $\delta g_{\bf k}=g_{\bf k}'({\bf K}=0)-g_{\bf k}$ and
$A(w_1,w_2)$ is a positive constant. In general, $\langle
w_1w_2|d_{\bf k}d^\dag_{\bf k}|w_1w_2\rangle=1-|g'_{\bf
k}(K=0)|^2$ is the quasihole distribution when the two vortices
are located at $w_1$ and $w_2$. Thus, $E_v$ is indeed the energy
cost to excite the vortex pair. Since $g'_{\bf k}\sim 1/k$ and
$E_{\bf k}\sim k$ for $k\to 0$, there is no infrared divergence in
$E_v$. Therefore it costs a finite energy for such vortex pair
excitations which is isolated to other states in the gapped B
phase. Note that $A(w_1,w_2)\to 0$ as $w_1$ and $w_2\to \infty$,
which corresponds to the case where the bulk of the system is sent
back to the vortex-free ground state.

\begin{figure}[htb]
\begin{center}
\includegraphics[width=4cm]{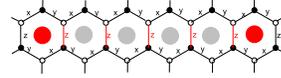}
\end{center}
 \caption{\label{fig:Fig. 4}
The vortex excitations. The grey solid circles denote $W_P=1$ and
the red solid circles denote vortices with $W_P=-1$. }
\end{figure}

The finiteness of $E_v$ and vanishing of $\langle \delta
H\rangle=0$ imply that the vortex excitations are isolate either
to the ground state or the other excitations. This is consistent
with analysis of Read and Green to the vortex excitations in U(1)
vortex excitation of the $p$-wave paired state \cite{rg}.

\end{document}